\documentclass[12pt]{article}
\usepackage[pdftex]{graphicx}
\usepackage{lineno,hyperref}
\usepackage{braket}
\usepackage[caption=false]{subfig}
\usepackage{amsmath}

\begin{document}

\title{Practical application of quantum neural network to materials informatics: prediction of the melting points of metal oxides}
\author{Hirotoshi Hirai\thanks{e-mail: hirotoshih@mosk.tytlabs.co.jp}\\
\\
\textit{Toyota Central R\&D Labs., Inc.,}\\
\textit{41-1, Yokomichi, Nagakute, Aichi 480-1192, Japan}}

\maketitle

\begin{abstract}
Quantum neural network (QNN) models have received increasing attention owing to their strong expressibility and resistance to overfitting.
It is particularly useful when the size of the training data is small, making it a good fit for materials informatics (MI) problems.
However, there are only a few examples of the application of QNN to multivariate regression models, and little is known about how these models are constructed.
This study aims to construct a QNN model to predict the melting points of metal oxides as an example of a multivariate regression task for the MI problem.
Different architectures (encoding methods and entangler arrangements) are explored to create an effective QNN model.
Shallow-depth ansatzs could achieve sufficient expressibility using sufficiently entangled circuits.
The ``linear'' entangler was adequate for providing the necessary entanglement.
The expressibility of the QNN model could be further improved by increasing the circuit width.
The generalization performance could also be improved, outperforming the classical NN model.
No overfitting was observed in the QNN models with a well-designed encoder.
These findings suggest that QNN can be a useful tool for MI.
\end{abstract}

\maketitle

\section{Introduction}

The application of machine learning (ML) to the development of materials is becoming increasingly important~\cite{butler2018machine, schmidt2019recent}.
Materials informatics (MI) is a field of information science used to develop materials~\cite{ramprasad2017machine, agrawal2016perspective, rajan2005materials}.
It involves constructing a predictive model of physical properties from a limited amount of data obtained from experiments or simulations and then screening materials with the desired performance from a large group of materials.
The challenge with MI is that the data are often limited and prone to noise owing to errors in the experimental data, making it difficult to construct a model with a good generalization performance (prediction performance for unknown materials)~\cite{butler2018machine,xu2023small}.

Recently, a quantum neural network (QNN)~\cite{abbas2021power}, also referred to as quantum circuit learning~\cite{mitarai2018quantum}, has been developed as an ML algorithm for quantum computers~\cite{steane1998quantum}.
It is a quantum-classical hybrid algorithm based on the variational quantum algorithm~\cite{cerezo2021variational}, which has been developed to work with noisy intermediate-scale quantum (NISQ) devices~\cite{preskill2018quantum}.
A QNN model is built by minimizing the discrepancy between the output of the quantum circuit and labeled data by adjusting the circuit parameters to their optimal values.
The advantage of QNN is that it can use high-dimensional quantum states as trial functions that are hard to generate on a classical computer~\cite{mitarai2018quantum}.
Another advantage of a QNN is that the unitarity of quantum circuits serves as regularization to prevent overfitting~\cite{mitarai2018quantum}.
In a classical neural network (NN) model, a regularization term is incorporated into the cost function to constrain the norm of the learning parameters and to reduce the model's expressibility to prevent overfitting~\cite{nielsen2015neural}.
In contrast, the norm of parameters is automatically limited to one due to unitarity in a QNN model, i.e., the regularization function is inherently provided.
QNNs have also been reported to afford predictive models with excellent generalization performance even when only a small amount of training data is available~\cite{caro2022generalization}.
It has also been reported that the smaller the data size of the problem, the greater the advantage of the generalization performance of QNNs over classical NNs~\cite{hirai2023application}.

These characteristics of QNNs may be particularly useful in MI.
The atomic configuration can be used to predict the properties of materials because the Hamiltonian can be determined from the atomic configuration and the Schrödinger equation can be solved (in principle) using the Hamiltonian to obtain the properties of the material.
 ML models can be used instead of solving the Schrödinger equation because solving the many-body Schrödinger equation is extremely difficult~\cite{brockherde2017bypassing}.
Such concepts have been considered in the MI~\cite{faber2017prediction} and QSAR (Quantitative Structure-Affinity Relationship)~\cite{tropsha2010best} fields.
The construction of an ML model that bypasses the Schrödinger equation is expected to be naturally aided by a QNN model with quantum architectures.

In this study, we attempted to construct a successful QNN model to predict the melting points of metal oxides.
Calculating thermodynamic properties such as melting points is difficult with first-principles calculations because of the high computational cost and lack of accuracy~\cite{sugino1995ab, puchala2013thermodynamics}.
Therefore, it is important to develop a practical melting point prediction model to identify functional materials~\cite{ward2016general, qu2019ultra}.
However, because QNNs are an emerging field, there is still a lack of understanding of how to construct effective QNN models.
We considered various architectures (ansatz and encoding methods) to create an effective QNN model for the practical task of predicting melting points.

\section{Methods}

\subsection{Data set}
This study addresses the issue of predicting the melting points of metal oxides.
The melting point data for metal oxides listed in~\cite{schneider1963compilation} were expanded to 70 metal oxides by adding data from other references~\cite{lide2004crc,coutures1989melting,wang2009pubchem}.
Each material was identified in the Materials Project database~\cite{jain2013commentary}, and the following five explanatory variables were obtained (some variables were calculated from structural data in the database in~\cite{jain2013commentary}).
\begin{itemize}
  \item formation\_energy\_per\_atom: Formation energy per atom
  \item band\_gap: Band gap energy
  \item density: Mass density
  \item cati\_anio\_ratio: Ratio of the number of cations and anions
  \item dist\_from\_o: Minimum distance from the oxygen atom to cation
\end{itemize}
The constructed dataset is available in the Supporting Information.
These explanatory variables were normalized to have a mean of 0 and a variance of 1 for the training data and further scaled to have a maximum value of 1 and a minimum value of -1.
The objective variable (melting point temperature in \textdegree C) was divided by 3500 and scaled such that the maximum value was approximately 1 (the highest melting point of metal oxides treated in this study was 3390 \textdegree C).

The k-fold cross-validation method~\cite{fushiki2011estimation} was used to evaluate the accuracy of the constructed regression models.
In this study, the 70 dataset entries were divided into five groups; one group was used as the test data, while the other groups were used as the training data.
This procedure was performed for all five combinations, and the average accuracy of the five models was used as the final accuracy.
The root mean square error (RMSE) was used as a measure of accuracy.

\subsection{QNN models}
The QNN model is composed of three components: an encoder that transforms explanatory variables into a quantum state, an ansatz which is a quantum circuit with learning parameters, and a decoder that converts the quantum state into an output value.
Each component is described in detail in the following sections.
In this study, QNN models were implemented using Pytket~\cite{sivarajah2020t}, a Python module for quantum computing, and quantum circuit calculations were performed using state vector calculations with the Qulacs~\cite{suzuki2021qulacs} backend, a quantum computing emulator.
The mean squared error (MSE) between the labeled data and model predictions was used as a cost function.
The Powell method~\cite{powell1964efficient} was used to optimize the learning parameters.

\subsubsection{Encoder}
\begin{figure}[ht]
\centering
\includegraphics[width=12cm]{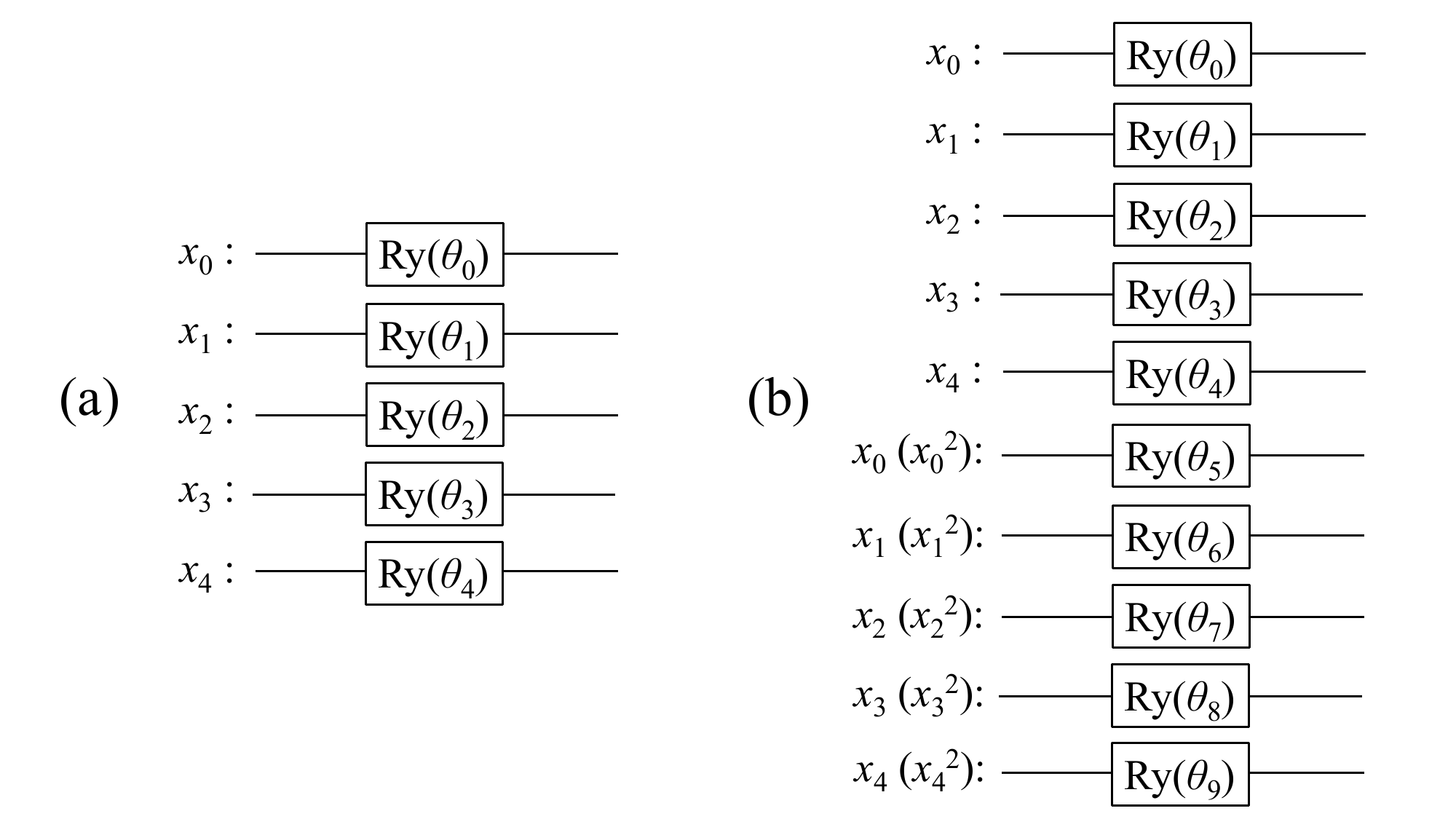}
\caption{The Ry encoders used in this study: (a) 5-qubit model and (b)10-qubit model.
The Ry gate acts on each qubit initialized to $\ket{0}$. The scaled explanatory variables $x_i$ (or $x^2_i$) are converted to the rotation angles $\theta_i$ according to $\theta = \arctan(x)+\pi/2$ or $\theta=\pi x$.}
\label{fig_encoder_setup}
\end{figure}

In this study, Ry rotation gates~\cite{nielsen2010quantum} were used as encoders.
We used two different methods to transform each scaled explanatory variable $x$ into the rotation angle $\theta$: $\theta = \pi x$ and $\theta = \arctan(x)+\pi/2$.
The arctangent allows the scaled explanatory variable to be uniquely converted to a rotation angle even if the value is outside the scale range (-1,1) when the scaler is used for the test data.
We constructed a 5-qubit QNN model with each explanatory variable encoded in one qubit and a 10-qubit QNN model with each explanatory variable encoded in two qubits, as shown in Fig.~\ref{fig_encoder_setup} (a) and (b), respectively.
In the 10-qubit model, two different encoding methods were tested:
one with redundant imputation of the explanatory variable $x$ and the other with imputation as $x$ and $x^2$, as indicated by the parentheses in Fig.~\ref{fig_encoder_setup} (b).

\subsubsection{Ansatz}
\begin{figure}[ht]
\centering
\includegraphics[width=12cm]{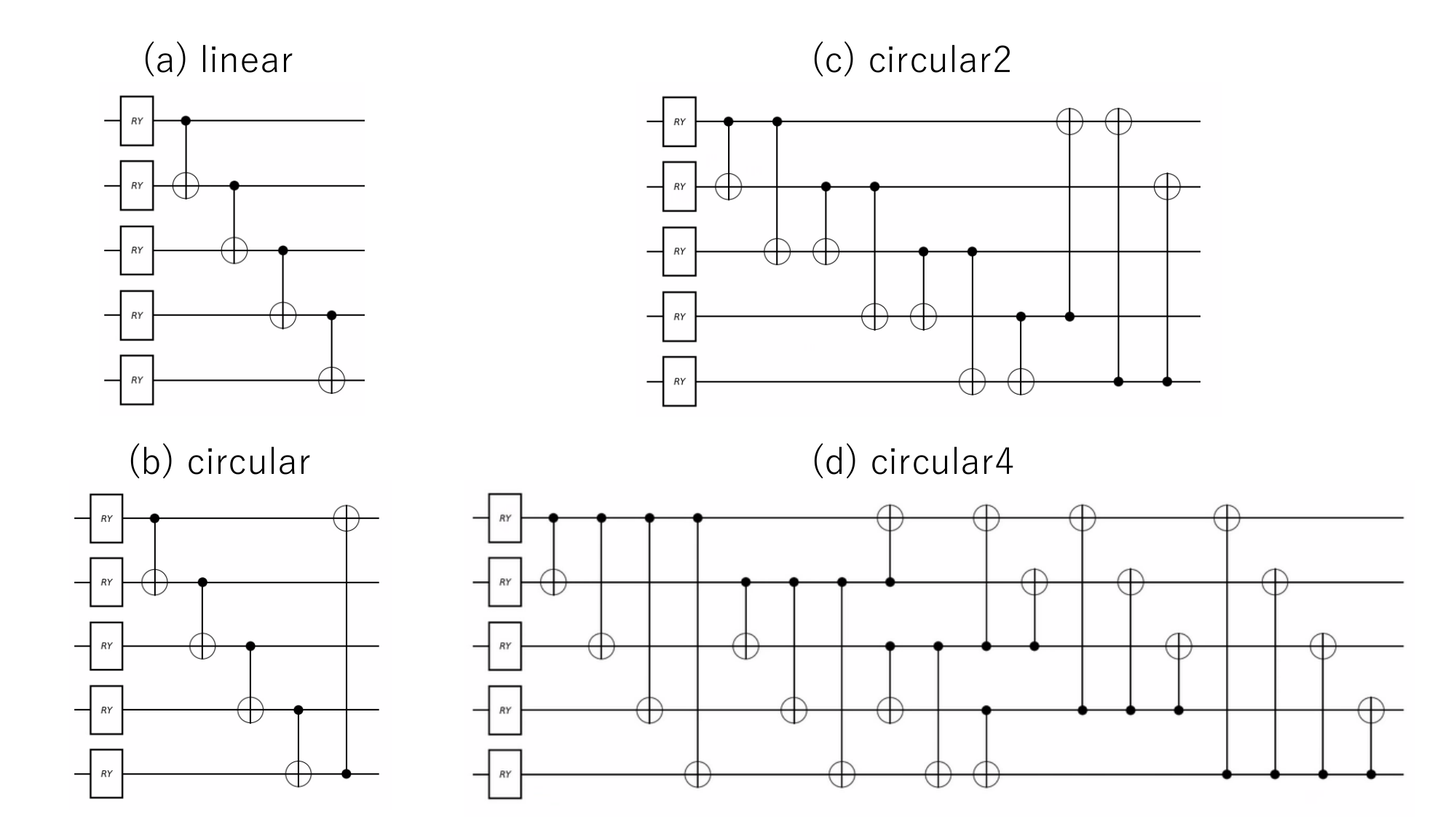}
\caption{The depth 1-block of each ansatz used in this study. These circuits consist of Ry rotating gates and entanglers (groups of 2-qubit operations).}
\label{fig_entangler_str}
\end{figure}

In this study, as the ansatz part of the QNN, we examined ansatzs with the quantum circuits shown in Fig.~\ref{fig_entangler_str} as the depth 1-block.
In these ansatzs, an entangler (a group of 2-qubit operations) was placed after the Ry rotation gate.
Although Fig.~\ref{fig_entangler_str} shows CNOT (CX) gates as 2-qubit gates, and we also examine the case using controlled-Z (CZ) gates.
circular2 (c) and circular4 (d) contain 2-qubit operations up to the second and fourth nearest-neighbor qubits, respectively.
Each Ry gate has an independent learning parameter $\theta$.
Because there are five (10) Ry gates in the depth 1-block of the 5-qubit (10-qubit) model, the number of parameters for the QNN model with depth $d$ is 5$d$ (10$d$).
In this study, $d$ values of 1 to 7 were considered.

\subsubsection{Decoder}
The QNN decoder takes the expectation value of an observable quantum state generated by the encoder-ansatz quantum circuit as the output of the regression model.
For the 5-qubit QNN models, the expectation value of $\sigma^4_z$ (the Z-axis projection of the lower-end qubit) was used as the decoder (note that the number on the label begins with zero).
For the 10-qubit QNN models, the expected value of $\sigma^4_z + \sigma^9_z$ was used.

\subsection{Circuit analysis}
The higher the expressibility of the ansatz, the better the regression accuracy.
Therefore, the quantitative evaluation of the expressibility of an ansatz plays an important role in the construction of a QNN model.
In this study, Kullback-Leibler (KL) divergence~\cite{nakaji2021expressibility} and entanglement entropy~\cite{sim2019expressibility} were used as ansatz evaluation tools.
In the KL divergence metric, the KL divergence between the fidelity distribution of quantum states obtained from an ansatz with random parameters and the fidelity distribution for Haar measurements is used to quantify expressibility~\cite{nakaji2021expressibility}. 
In the entanglement entropy, the entanglement entropy between one qubit and another was calculated for an ansatz with random parameters and the statistical mean obtained.
This calculation was performed for different qubits as subsystems, and the average of the qubits was used to quantify the entanglement strength of the ansatz.

\subsection{Classical NN models}
A conventional neural network (NN) model was constructed for comparison.
To vary the number of learning parameters in the NN regression model, models 5-5-1(36), 5-3-1(22), 5-2-1(15), 5-1(6) were prepared, where the numbers indicate the number of neurons in the fully connected layers, ``-'' indicates ``between layers'', and the numbers in parentheses represent the number of training parameters. 
A sigmoid function was used as the activation function.
PyTorch~\cite{imambi2021pytorch} is used to construct and train the NN model.
The Adam optimizer~\cite{kingma2014adam}, an extended version of the stochastic gradient descent, was used with a learning rate of 0.02 over 10000 epochs. 
L2 regularization was applied to prevent overfitting.
The weight parameter for L2 regularization (a hyperparameter set by the user) was used to minimize the RMSE for the test data (average of five groups).
We tested the parameters of $10^{-n}$ with $n =$ 2, 3, 4, and 5, and found that $n = 4$ gave the best performance for all models.

\section{Results and discussions}

\subsection{Encoder}
\begin{figure}[ht]
\centering
\includegraphics[width=12cm]{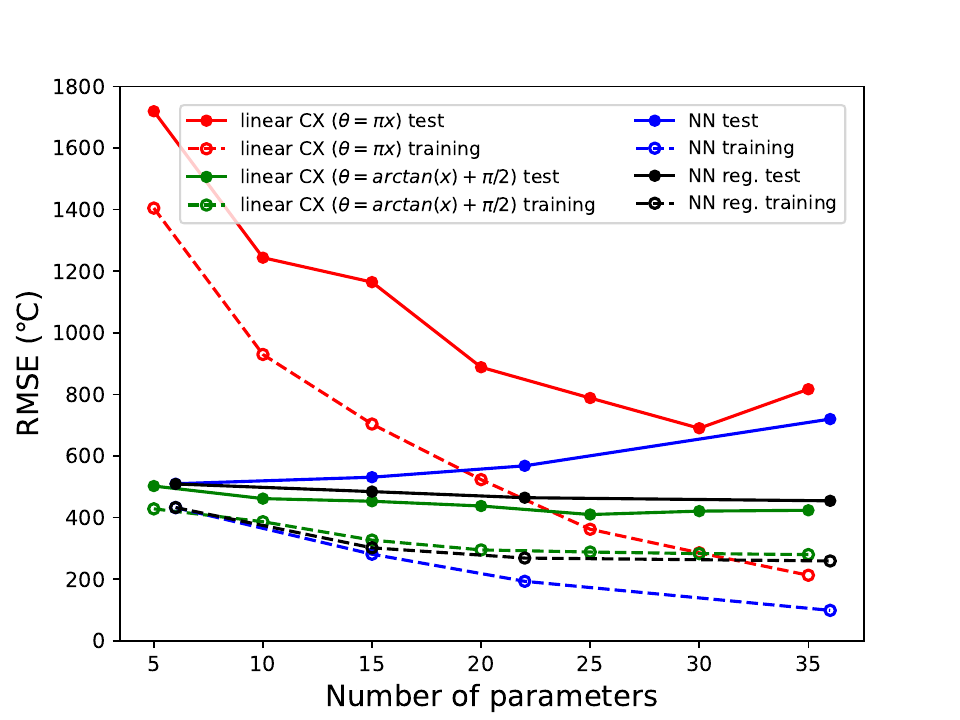}
\caption{The RMSE for the QNN models with Ry($\pi x$) and Ry($\arctan (x)+\pi /2$) encoders. The number of qubits is fixed to five and the entangler is fixed to the linear arrangement. The classical NN results with and without regularization are also shown.}
\label{fig_encoder}
\end{figure}

First, we present the results of the analysis of the effects of different methods on transforming the explanatory variable $x$ into the rotation angle $\theta$ during Ry($\theta$) encoding.
The RMSE of the QNN models with Ry($\pi x$) and Ry(arctan($x$)+$\pi$/2) are shown in Fig.~\ref{fig_encoder}.
Here, the number of qubits was fixed at five, and the entangler was fixed in a linear arrangement (Fig.~\ref{fig_entangler_str} (a)).
The number of parameters in the model increased with the depth of the ansatz.
For comparison, Fig.~\ref{fig_encoder} also shows the results for the classical NNs with and without regularization as ``NN reg.'' and ``NN'', respectively.
It can be confirmed that NN models without regularization induce overfitting.
That is, the RMSE of the test data increases as the number of parameters increases.
When Ry($\pi x$) was used as the encoder, QNN models with a small number of parameters (shallow ansatzs) exhibited significantly poorer regression performance.
The reasons for this are as follows.
Here, the explanatory variable $x$(-1,1) is converted into a rotation angle $\theta$(-$\pi$,$\pi$), which results in a round trip around the Bloch sphere, and the Z-axis projection after encoding is not unique.
In extreme cases, $x=-1$ and $x=1$ are encoded in the same quantum state.
As the number of parameters increases (the ansatz is deepened), the RMSE becomes smaller for the training data.
This is thought to be because the data are fully trained by brute force with a large number of parameters. 
However, for the test data, overfitting was observed for the models with deep ansatzs.
However, in the QNN model using Ry(arctan($x$)+$\pi$/2) as the encoder, the RMSE was small, even for a model with a small number of parameters (shallow ansatzs).
It can also be confirmed that overfitting does not occur even in models with a large number of parameters (deep ansatzs).
In this case, the RMSE values for the test and training data showed approximately the same dependence on the number of parameters as the classical NN with regularization, confirming that the automatic regularization function of the QNN was effective.
In the following discussion, Ry(arctan($x$)+$\pi$/2) was used as the encoder.

\subsection{Ansatz}
\begin{figure}[ht]
\centering
\includegraphics[width=12cm]{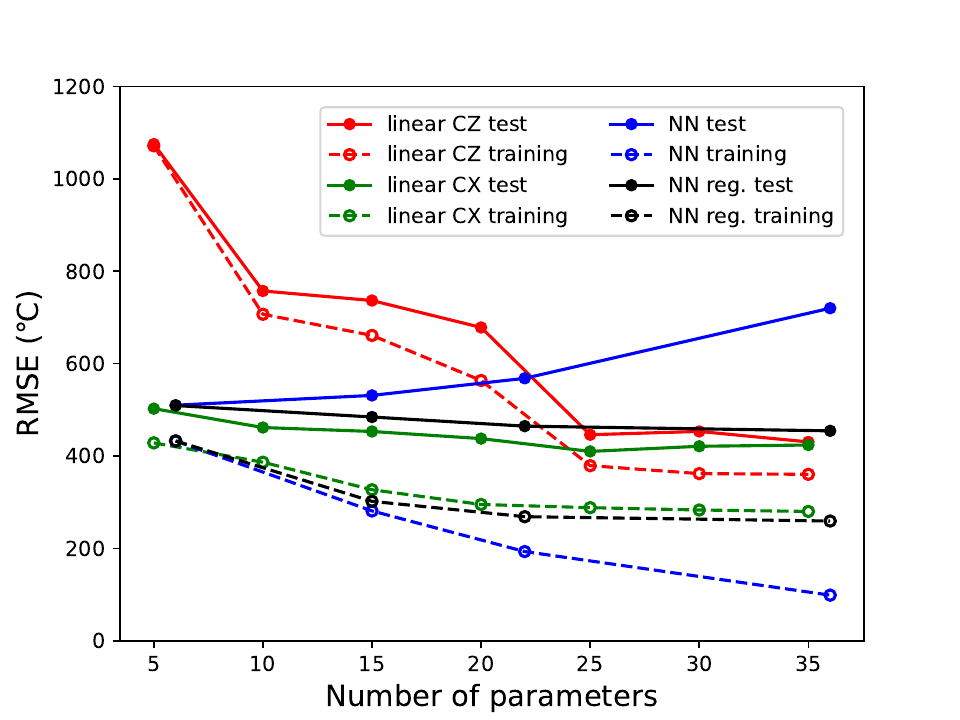}
\caption{The difference between the CX and CZ gates for QNN regression model performance. The number of qubits is fixed to five and the entangler is fixed to the linear arrangement.}
\label{fig_cz_cx}
\end{figure}

Next, we analyzed the impact of ansatz differences on the regression performance of the QNN.
The differences between the CX and CZ gates is shown in Fig.~\ref{fig_cz_cx}, where the number of qubits is fixed to five and the entangler is fixed to the ``linear'' arrangement.
From the comparison of the ansatzs with the CX and CZ gates, the QNN models with CZ have lower expressibility.
Because the observation axis is set to the Z axis ($\sigma_z$ is used for the decoder), the phase inversion by the CZ gate does not directly change the projection of the Z axis (the Pauli gate based on the basis axis does not change the state, except for the phase).
As a result, QNN models with CZ gates are considered to have lower expressibility, particularly when the number of parameters is small.
In the following discussion, only CX gates were used as entanglers.

\begin{figure}[ht]
\centering
\includegraphics[width=12cm]{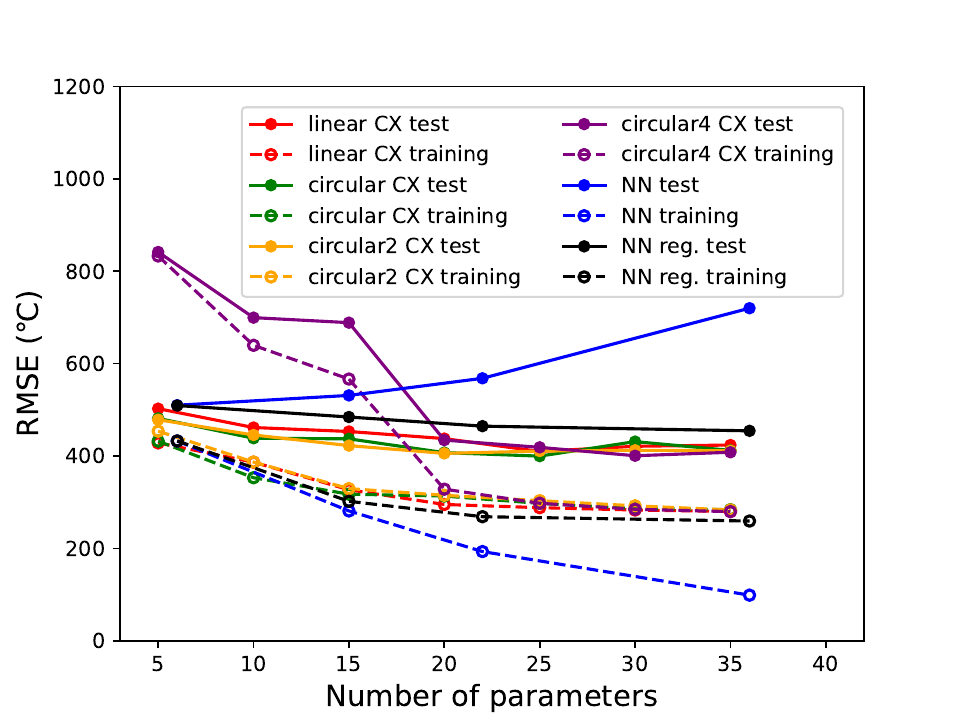}
\caption{The impact of different entangler structures on QNN performance.}
\label{fig_entangler}
\end{figure}

\begin{figure}[ht]
\centering
\includegraphics[width=12cm]{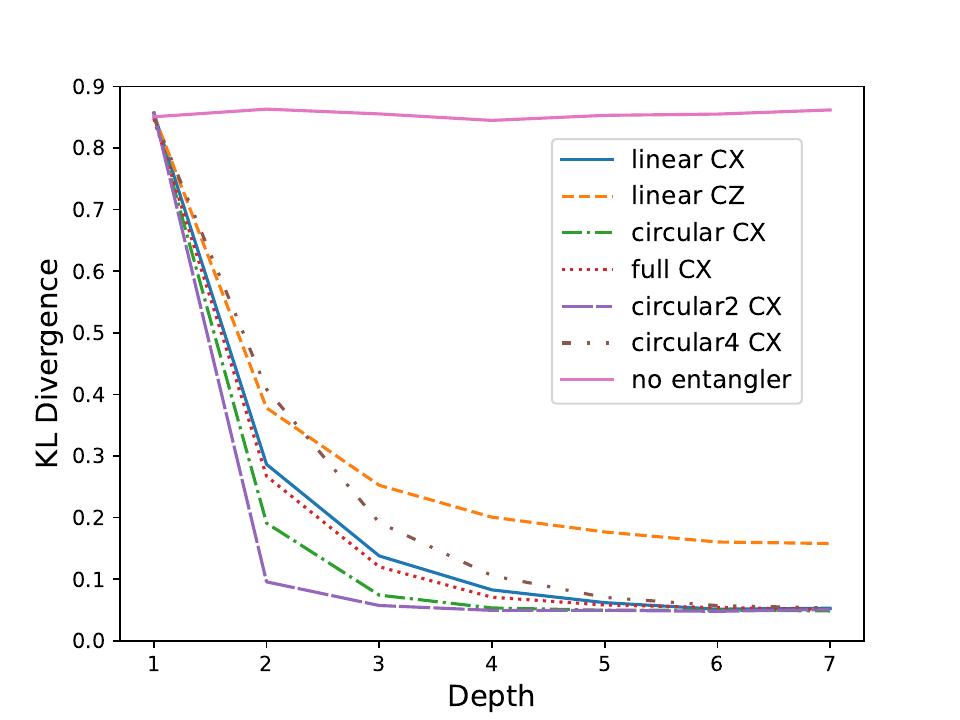}
\caption{The KL divergence of each ansatz.}
\label{fig_kld}
\end{figure}

\begin{figure}[ht]
\centering
\includegraphics[width=12cm]{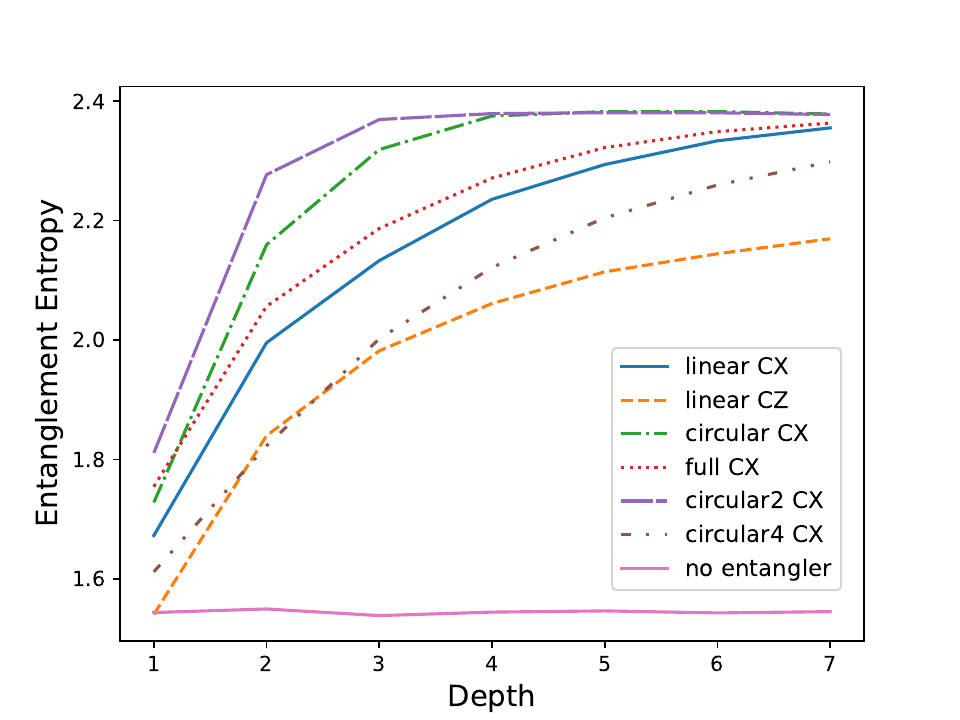}
\caption{The entanglement entropy of each ansatz.}
\label{fig_entropy}
\end{figure}

\begin{figure}[ht]
\centering
\includegraphics[width=12cm]{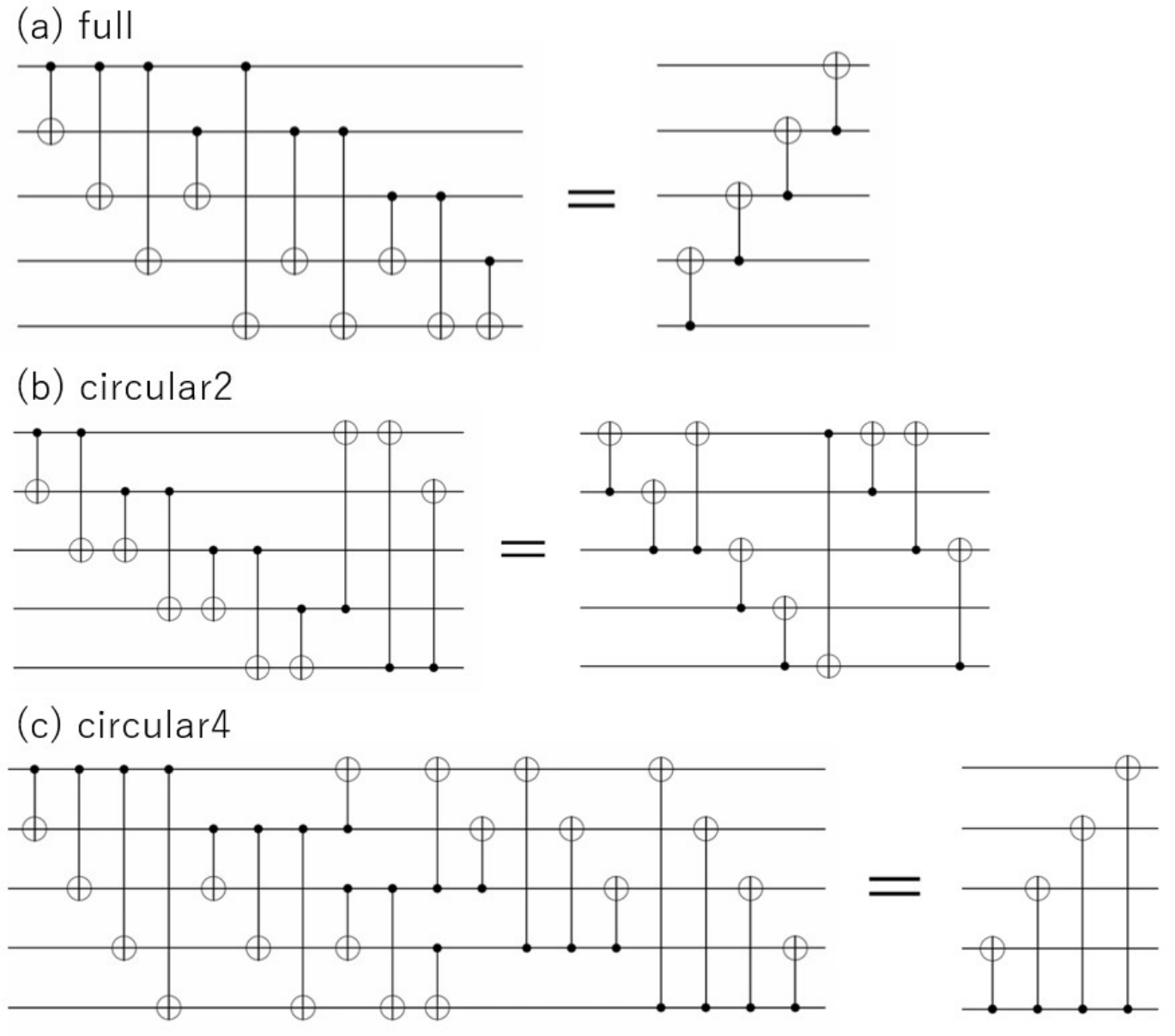}
\caption{Each entangler and its equivalent reduced circuit.}
\label{fig_touka}
\end{figure}

Fig.~\ref{fig_entangler} shows the impact of different entangler structures (Fig.~\ref{fig_entangler_str}) on QNN performance.
The QNN models with ansatz ``linear'', ``circular'', and ``circular2'' show similar performances, while the QNN model with ansatz ``circular4'' performs significantly worse for a small number of parameters (shallower depths).
To investigate the factors contributing to these results, the KL divergences and entanglement entropies of these ansatzs were examined, as shown in Fig.~\ref{fig_kld} and Fig.~\ref{fig_entropy}, respectively.
These figures also show the results for the ``full'' arrangement shown in Fig.~\ref{fig_touka} (a).
These results indicate a correlation between KL divergence and entanglement entropy, with a larger entanglement entropy indicating a smaller KL divergence.
Therefore, an ansatz with larger entanglement has greater expressibility.
It can be expected that the entanglement becomes stronger as the number of CXs increases, such as ``linear'', ``circular'' and ``circular2'', but it is noticeably weaker for the ``full'' and ``circular4'' entanglers.
This can be understood based on the following facts:
It is known that a ``full'' entangler has a reduced circuit and is equivalent to an inverse ``linear'' entangler~\cite{ballarin2023entanglement} (Fig.~\ref{fig_touka} (a)).
This implies that entanglement cannot be enhanced by blindly including a large number of CXs, provided that a simple equivalent circuit (reduced circuit) exists.
However, it is difficult to determine whether a circuit has a reduced equivalent circuit.
Therefore, we optimized each entangler using the circuit optimization function in tket~\cite{sivarajah2020t} and explored a reduced equivalent circuit.
The results are summarized in Fig.~\ref{fig_touka}.
There is a significantly reduced equivalent circuit for ``circular4''.
In the reduced ``circular4'' entangler, each qubit has only a CX gate with the bottom qubit, so the entanglement is weak, as can also be seen from the entanglement entropy.
In contrast, ``circular2'' is not significantly simplified, and the entanglement is not notably weak.

Figures~\ref{fig_kld} and \ref{fig_entropy} show the KL divergence and the entanglement entropy for the ``linear CZ'' entangler, respectively, and indicate that the QNN model with this entangler has less expressibility.
These results indicate that KL divergence and entanglement entropy may be able to screen out ansatz with poor expressibility.

In this study, there were no large differences in QNN performance among ansatzs with entanglement greater than the ``linear'' entangler, and therefore, the ``linear'' entangler was found to provide sufficient entanglement for the QNN model for this problem.
This implies that a model with satisfactory performance can be constructed using only 2-qubit operations between neighboring qubits, suggesting that it may be feasible to operate the QNN model on superconducting quantum computers, which are widely used today, in the near future.

\subsection{Circuit width}
\begin{figure}[ht]
\centering
\includegraphics[width=12cm]{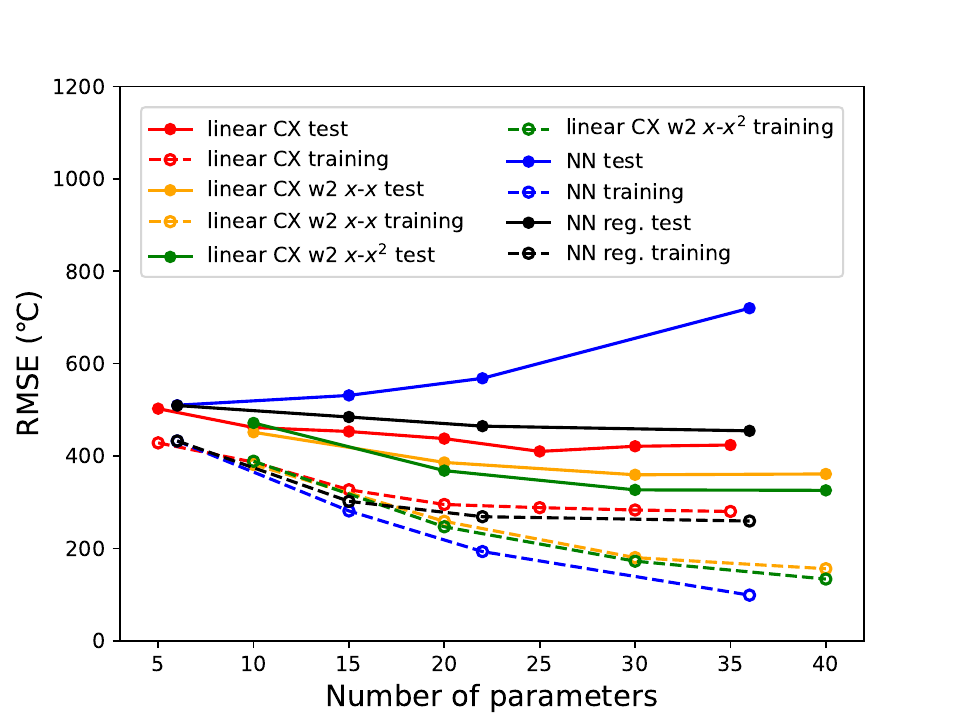}
\caption{The effect of the number of qubits (circuit width) on the performance of the QNN model.}
\label{fig_w1_w2}
\end{figure}

The effect of the number of qubits (circuit width) on the performance of the QNN model is illustrated in Fig.~\ref{fig_w1_w2}.
Here, the entangler is fixed to the ``linear'' arrangement.
When comparing the RMSE for the training data, the model with twice the number of qubits (w2) had a smaller error than the original model, indicating that its expressibility was improved by increasing the basis dimension.
The generalization performance (accuracy for the test data) was also improved by increasing the circuit width and outperformed the classical NN model.
Comparing the model with redundant inputs of the explanatory variable ($x$-$x$) and the model with redundant inputs ($x$-$x^2$), the latter appears to perform slightly better.
This is because it prevents basis duplication and efficiently handles a large number of basis functions.

\section{Conclusion}
In this study, we constructed QNN models to predict the melting point of metal oxides by exploring various architectures (encoding methods and entangler arrangements).
The explanatory variables should be uniquely converted into rotation angles to obtain good QNN models and avoid overfitting.
It was also found that even shallow-depth ansatzs could achieve sufficient expressibility for the present task using sufficiently entangled circuits.
It is insufficient to place a large number of CX gates without consideration; it is necessary to set up an entangler that produces entangles in real terms.
In this case, KL divergence and entanglement entropy proved to be good indicators.
The ``linear'' entangler was adequate for providing the necessary entanglement for the QNN model for this particular problem.
This result indicates that a model with satisfactory performance can be created using only 2-qubit operations between adjacent qubits.
The expressibility of a QNN model can be improved by increasing the circuit width (number of qubits).
This also improved the generalization performance, outperforming the classical NN model.
Most importantly, no overfitting was observed in QNN models with well-designed encoders.
A QNN can achieve high generalization performance without hyperparameter tuning and is considered an excellent tool for regression tasks.

\section*{Conflicts of interest}
The authors declare no conflict of interest.

\section*{Supporting Information}
The melting point data for the metal oxides and the explanatory variables used in this study are listed in the following.
\begin{figure}[ht]
\centering
\includegraphics[width=12cm]{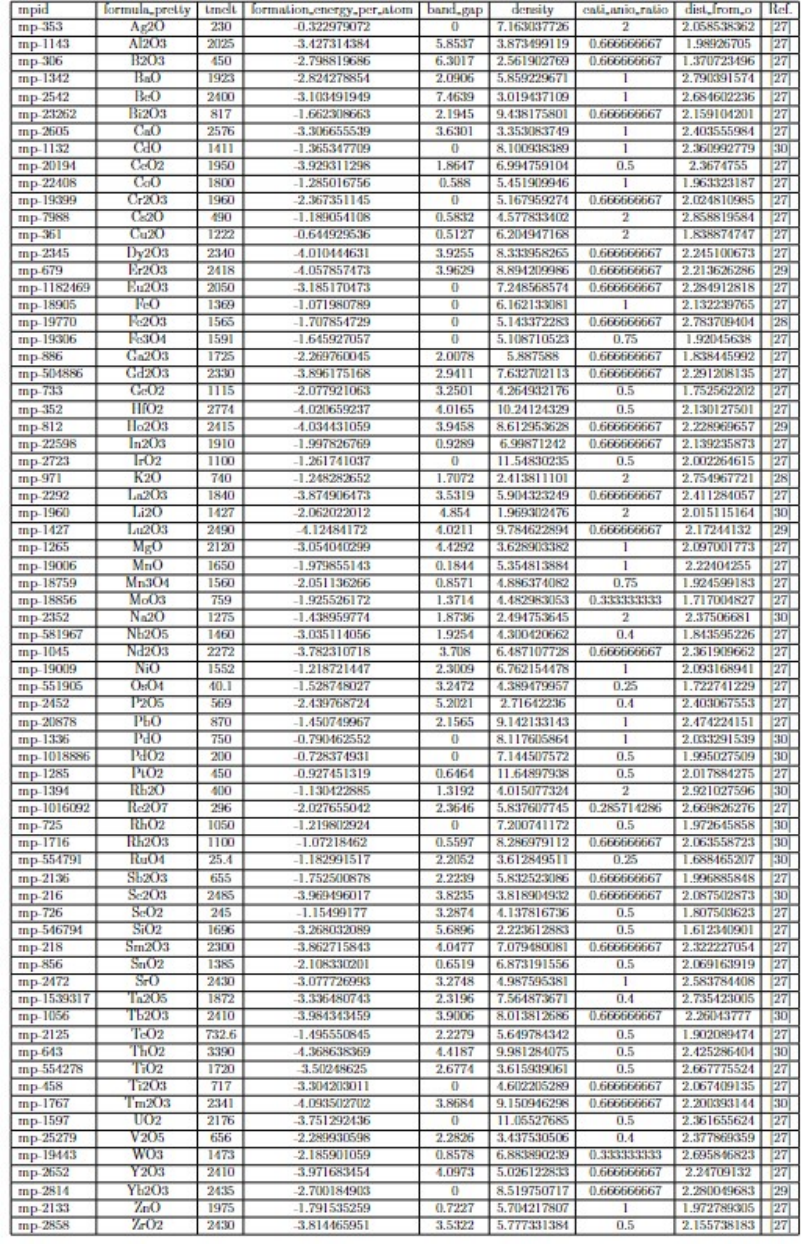}
\end{figure}

\clearpage
\pagebreak
\bibliography{ref}

\end{document}